\documentclass[journal]{IEEEtran}
\usepackage{amsmath}
\usepackage{amsfonts,amssymb}
\usepackage{graphicx}
\allowdisplaybreaks[4]
\usepackage{xcolor}
\definecolor{deepred}{RGB}{205,38,38}
\usepackage{algorithm}
\usepackage{algorithmic}
\usepackage{amsthm}

\usepackage{xpatch}
\usepackage{lineno,hyperref}
\modulolinenumbers[5]
\usepackage{bm}

\usepackage{bbm}
\usepackage{multirow}
\usepackage{makecell}
\usepackage{array}
\usepackage{booktabs}
\usepackage{lipsum}
\ifCLASSINFOpdf

\else

\fi

\ifCLASSOPTIONcompsoc
  \usepackage[caption=false,font=normalsize,labelfont=sf,textfont=sf]{subfig}
\else
  \usepackage[caption=false,font=footnotesize]{subfig}
\fi

\begin{document}

\title{Two-Layer Volt/VAR Control in Unbalanced Active Distribution Systems: Efficient Optimization \\and Accurate Tracking}

\author{Yifei Guo,~\IEEEmembership{Member,~IEEE,}
Qianzhi Zhang,~\IEEEmembership{Student Member,~IEEE,}
Zhaoyu Wang,~\IEEEmembership{Member,~IEEE}\\
Fankun Bu,~\IEEEmembership{Student Member,~IEEE,}
and Yuxuan Yuan,~\IEEEmembership{Student Member,~IEEE,}
\thanks{This work was supported by the U.S. Department of Energy Wind Energy Technology Office under DE-EE8956 and the National Science Foundation under ECCS 1929975. \emph{(Corresponding author: Zhaoyu Wang)}}
\thanks{The authors are with the Department of Electrical and Computer Engineering, Iowa State University, Ames, IA 50011 USA. (email:yifeig@iastate.edu; qianzhi@iastate.edu;wzy@iastate.edu; fbu@iastate.edu; yuanyx@iastate.edu)}% <-this % stops a space
}

\markboth{Submitted to IEEE for possible publication. Copyright may be transferred without notice}%
{Shell \MakeLowercase{\textit{et al.}}: Bare Demo of IEEEtran.cls for Journals}

\maketitle

\begin{abstract}
This paper proposes a novel two-layer Volt/VAR control (VVC) framework  to regulate the voltage profiles across an unbalanced active distribution system, which achieves both the efficient open-loop optimization and accurate closed-loop tracking. In the upper layer, the conventional voltage regulation devices with discrete and slow-response characteristics are optimally scheduled to regulate voltage profiles in an hourly timescale while improving economic operations based on the receding horizon optimization (RHO) in a centralized  manner. A generalized linearized branch flow model (G-LBFM) is developed to incorporate tap changers into branches, which significantly reduces the computational complexity compared to the original mixed-integer non-convex case. In the lower layer, we develop an integral-like control algorithm rather than resorting to the droop-based rules for real-time reactive power dispatch of distributed energy resources (DERs) to achieve accurate voltage tracking  and mitigate fast voltage fluctuations in a decentralized (purely local) fashion. Further, a sufficient stability condition of the integral rule is presented to guarantee the closed-loop stability. Case studies are carried out on the unbalanced IEEE 123-Node Test Feeder to validate the effectiveness of the proposed method.
\end{abstract}


\IEEEpeerreviewmaketitle

\section{Introduction}
\IEEEPARstart{V}{oltage}/reactive power (VAR) control (VVC) is an essential task to ensure the secure operation of distribution systems. Generally, it is performed by using a predefined set of rules to schedule the various conventional voltage regulation devices, including capacitor banks (CBs), step-voltage regulators (SVRs) and on-load tap changing transformers (OLTCs), which might not able to tackle fast voltage issues caused by the high variability of loads and rapidly developed distributed energy resources (DERs), e.g., distributed wind and photovoltaic (PV) generation, due to their slow response and limited operation  times.  
Therefore, the inverter-based DERs with fast and continuous VAR capability have been encouraged to provide necessary Volt/VAR support for distribution systems \cite{IEEE1547_2018}. %As advocated by the IEEE Std. 1547-2018 , a linear droop control strategy of inverter-based DERs could be a feasible choice in practice, which only relies on the local voltage feedback.  %\cite{droop_VVC_1} and \cite{droop_VVC_2}, local droop controllers for multiple inverters have been designed to absorb or generate reactive power for mitigating real-time voltage deviations and achieving plug-and-play VVC. 
%However, because of the inherent proportional control design and the high sensitivity of the control parameters, the limitations of droop control based VVC are obvious, such as the vulnerability of instability and inaccurate set-point tracking \cite{VVC_review}. 

To optimally coordinate overall Volt/VAR regulation resources and achieve specific control goals, such as loss reduction and voltage deviation mitigation, the centralized or distributed optimization methods have been widely investigated \cite{Pal}--\cite{Feng}. See \cite{OV1} and \cite{OV3} for related surveys. Given such optimization-based methods are mostly designed in open-loop fashion, potential model errors, prediction errors and communication delays/noises would  deteriorate their performance for real-time implementations \cite{haozhu}. Besides, the 
\emph{single-layer} optimization/control might not be able to exploit different response capabilities of various VVC devices in the best way.

In this context, the idea of \emph{multi-layer} (also termed as \emph{multi-level} or \emph{multi-stage}) control has attracted a lot of attention, which aims to coordinate different devices in separable timescales.
In \cite{LM_two_stage}--\cite{JYP_two_stage}, the coordination between  OLTC and CBs in traditional distribution systems were coordinated using two-stage frameworks.
In \cite{DJ_multi_time}, a two-timescale Volt/VAR optimization method was proposed wherein the OLTCs and CBs are dispatched in hourly timescale while the DERs are dispatched in 15-min timescale explicitly considering their uncertainties. Similarly in \cite{Yifei2}, a two-stage open-loop optimization model based on model predictive control was established to coordinate multiple devices. In \cite{Yu}, a bi-level voltage management scheme was proposed, including the centralized coordination of OLTCs and CBs and a distributed consensus-based algorithm to dispatch electric springs working on a faster timescale as supplementary control for critical loads. In \cite{YWang}, a distributed two-layer VVC scheme was developed to dispatch the grouped PV inverters, including a distributed optimization-based 15-min dispatching and droop-based real-time control.  The authors in \cite{HSB}--\cite{THong_droop} proposed to schedule the parameters of droop controllers by upper-layer optimization.
In \cite{Czhang_multi_time}, the robust VVC was specially addressed.

It can be concluded that a preferable design of the multi-layer VVC is the combination of open-loop optimization to schedule the discrete devices with slow response and an easy-to-implement control law with minimal computation and communication requirements in real-time layer to dispatch the flexible DERs. However, due to existence of discrete devices and nonlinearity of ac power flow, the  optimization problems are essentially non-convex mixed-integer nonlinear programming problems which become intractable for large systems. To tackle this complexity issue, some heuristics methods such as particle swarm optimization \cite{LM_two_stage,DJ_multi_time} and differential evolution \cite{Yu}, are used to solve the problems or some standard nonlinear programming algorithms are adopted after relaxing the discrete variables to continuous ones \cite{Pal,Anna,BAR_dis_tap}. However, these methods are sometimes time-consuming and often suffer the sub-optimality \cite{Yifei2}.  The three-phase unbalanced cases may further complicate the optimization problems, which was not addressed in most previous multi-layer architectures.
For the real-time layer, as advocated by \cite{IEEE1547_2018}, a popular choice is the linear (piece-wise) droop control \cite{YWang}--\cite{THong_droop}, which, however, may be challenged by its vulnerability to closed-loop instability due to improper control parameter selection and inaccurate voltage tracking due to its inherent proportional rule \cite{OV1}. 
%However, a simplified single phase power flow model is applied in \cite{Czhang_multi_time}, which neglects the unbalanced features and the coupling between phases in distribution system. As mentioned before, the local  droop controllers adopted in \cite{Czhang_multi_time} also suffers various disadvantages when doing the real-time control of DERs. Thus, to address the shortcomings in the existing VVC methods, an efficient multi-layer framework for coordinated voltage regulation devices and an accurate real-time control method for DERs are necessary.     

In this paper, we offer a two-layer VVC framework to coordinate the optimal settings of OLTC, SVRs and CBs and the real-time VAR adjustment of inverter-based DERs in separable timescales. The upper-layer control is developed based on the centralized receding horizon optimization (RHO) with the help of rolling predictions of DER generation and load consumption, which is established using the three-phase generalized linearized branch flow model (G-LBFM) that incorporates a tap changer over a branch. In the lower layer, we propose an integral-like algorithm with the aim of tracking the voltage references scheduled by the upper layer control, instead of resorting to the droop control to adjust the VAR outputs of DER inverters. Compared with existing methods, the advantages of the proposed method can be summarized as follows:
\begin{itemize}
\item With the G-LBFM, the nonlinearity of three-phase power flow  caused by tap changers can be approximated by a linear model with a sufficient accuracy. In this way, the RHO model becomes a standard mix-integer quadratic programming (MIQP) problem that can be efficiently handled by solvers. 
\item Compared with the droop control, the proposed integral-like control can achieve more accurate voltage tracking by fully exploiting the VAR capabilities of inverters. Due to closed-loop nature of proposed method, better tracking performance has been shown compared with the optimization method based on  system-wide information. 
\item A sufficient condition with rigorous proof is provided to avoid Volt/VAR hunting/oscillation and ensure the closed-loop stability of the lower-layer control.
\end{itemize}

The reminder of the paper is organized as follows. Section \ref{sec:overview} gives a overview of the proposed two-layer method. Section \ref{sec:model} presents the generalized branch flow and voltage regulation device models. Section \ref{sec:stage1} present the open-loop upper-layer control. Section \ref{sec:stage2} presents the decentralized real-time VAR control algorithm of DERs in the lower layer. Simulation results are given in Section \ref{sec:case}, followed by conclusions.

\section{Overview of Two-Layer Volt/VAR Control}\label{sec:overview}
The basic idea of the two-layer VVC is to deal with different voltage issues in separable timescales (voltage deviations in the timescale of a few hours and voltage fluctuations in the timescales ranging from a few seconds to minutes) by exploiting multiple VVC resources with different response time.

The upper layer aims at optimally scheduling the discrete devices OLTC, SVRs and CBs based on the receding horizon optimization to improve the economic operation of the distribution systems while regulating the voltage profile within the predefined range $[0.95,1.05]$ p.u.. 
The RHO method is used here to address the operation limits of switchable devices over a given period and more accurate prediction of load and DER generation (wind/solar). It is inherently an open-loop method that requires some prior information (e.g., network topology and parameters) and some online operation information (e.g., wind, solar and load prediction and operation status of OLTC, SVRs and CBs). Note that, the potential VAR capabilities of DERs are considered in the optimization rather than being neglected to avoid overuse of other devices 
The corresponding optimal voltage solution of the RHO problem will be used as references for the lower layer control, resulting in a hierarchical control structure.

The lower layer aims to mitigate the voltage fluctuations in the timescales ranging from seconds to minutes by commanding the VAR outputs of DER inverters. An integral-like real-time control algorithm is developed to track the voltage references scheduled by the upper layer. Each DER inverter will update its VAR output according to its local voltage measurements. Given the algorithm requires very little computation and local information, it can be carried out in the time scale of a few seconds in a decentralized manner.The closed-loop nature of lower-layer control can largely reduce the risk of voltage violations, which might be caused by the model or prediction errors in the upper-layer optimization. 

\section{System Models}\label{sec:model}
\subsection{Generalized Branch Flow Model}
In this subsection, we propose a \emph{generalized branch flow model} to take into account the existence of SVRs, which is not addressed in the traditional BFM. Additionally, it will be extended to unbalanced three-phase systems.
\subsubsection{Single-Phase Model}
A distribution system is typically operated with tree topology. Consider a radial network comprising $N+1$ buses denoted by set $\mathcal{N}\bigcup\{0\}$, $\mathcal{N}:=\left\{1,\ldots,N\right\}$ and $N$ branches denoted by $\mathcal{E}$. Bus 0 denotes the slack bus (high-voltage side of main transformer). For each bus $j\in\mathcal{N}$, $p_{{\rm c},j}$ and $q_{{\rm c},j}$ are the real and reactive power consumptions; $p_{{\rm inv},j}$ and $q_{{\rm inv},j}$ are the real and reactive power injections by the DER inverter; $q_{{\rm cap},j}$ denotes the VAR injections from capacitor banks; $v_{j}:=|V_{j}|^2$ represents the squared voltage magnitude; $\mathcal{N}_{j}$ denotes the set of children buses. For any branch $({ i,j})\in\mathcal{E}$, $r_{ij}$ and $x_{ij}$ are the branch resistance and reactance while $P_{ij}$ and $Q_{ij}$ denote the real and reactive power flow from bus $i$ to bus $j$, respectively.
\begin{figure}[t]
  \centering
  \includegraphics[width=2.5in]{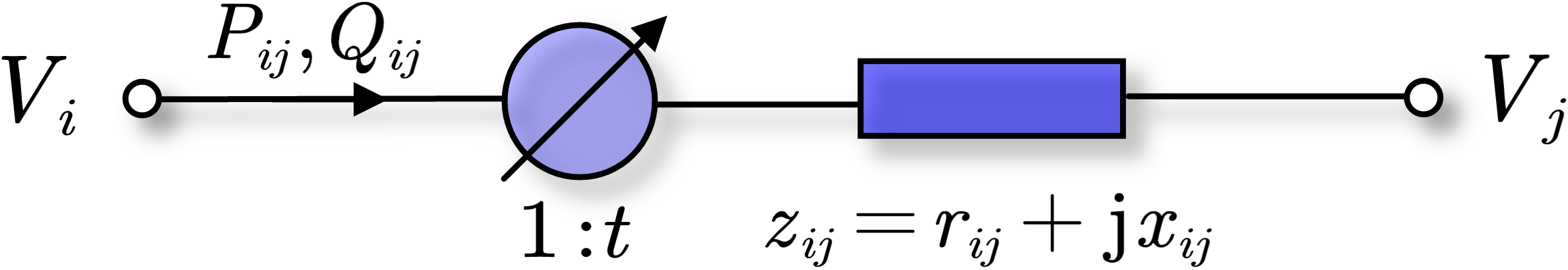}\\
  \caption{Generalized single-phase branch model.}\label{1ph_branch}
\end{figure}

A generalized single-phase branch model with a tap changer is shown in Fig. \ref{1ph_branch}. Upon this, the LBFM is generalized to, $\forall (i,j)\in\mathcal{E}$,
\begin{subequations}\label{BFM_alpha}
\begin{align}
P_{ij}&=\sum_{k\in\mathcal{N}_{j}} P_{jk}+p_{{\rm c},j}-p_{{\rm inv},j}\\
Q_{ij}&=\sum_{k\in\mathcal{N}_j} Q_{jk}+q_{{\rm c},j}-q_{{\rm inv},j}-q_{{\rm cap},j}\\
t_{ij}^2\cdot v_{i}-v_{j}&=2(r_{ij}P_{ij}+x_{ij}Q_{ij})\\[1mm]
t_{ij}&=1+n_{{\rm tap},ij}\cdot\Delta{tap}_{ij}
\end{align}
\end{subequations}
where $n_{{\rm tap},ij}$ and $\Delta{tap}_{ij}$ denote the tap position and step, respectively. Generally, the series impedance and shunt admittance of SVR (auto-transformer) are small in per unit, so that they can be neglected here \cite{Kersting}. %It can be also aggregated into the series impedance $z$ for more accurate analysis.

\begin{figure}[t]
  \centering
  \includegraphics[width=2.4in]{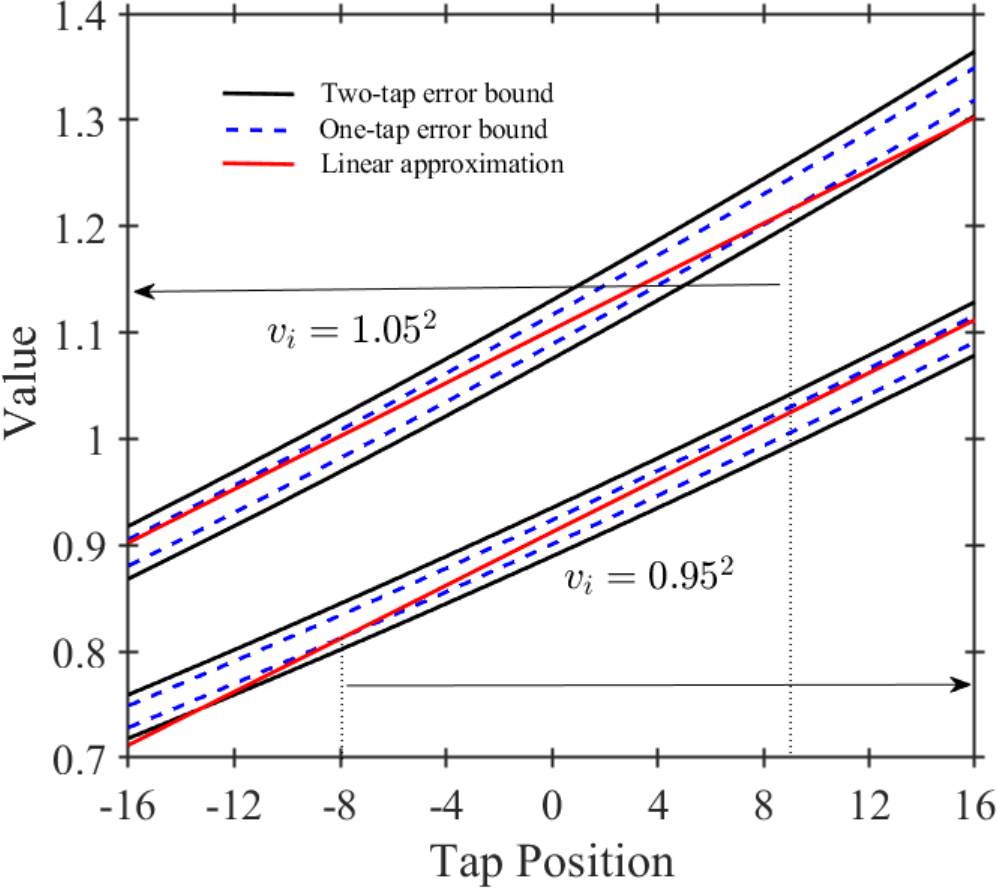}\\
  \caption{Model error of a linear approximation of term $t_{ij}^2\cdot v_{i}$.}\label{error}
\end{figure}

Obviously, the existence of $t_{ij}^2v_i$ makes the optimization-based VVC problems become nonlinear nonconvex mixed-integer problems. As known, the constraints with integer variables can be handled by mixed-integer solvers using the algorithms such as branch-and-bound, provided its relaxation is convex. However, this is not the case with (\ref{BFM_alpha}c) because of  the multiplication of $t_{ij}^2v_i$ and the quadratic term $t_{ij}^2$. That is, even after relaxing the integer variables, the constraints still hold a nonlinear nature, making the problem hard to solve. To tackle with this, a linear approximation of $t_{ij}^2\cdot v_{i}$ is derived as,
\begin{align}\label{linearization}
\nonumber t_{ij}^2\cdot v_{i}&=\left(1+2n_{{\rm tap},ij}\cdot\Delta{tap}_{ij}+n_{{\rm tap},ij}^2\cdot\left(\Delta{tap}_{ij}\right)^2\right)\cdot v_{i}\\
\nonumber &\approx v_{i}+2n_{{\rm tap},ij}\cdot\Delta{tap}_{ij}\cdot v_{i}\\
&\approx v_{i}+2n_{{\rm tap},ij}\cdot\Delta{tap}_{ij}\cdot v_{{\rm nom}}.
\end{align}
Such an approximation is believed to hold because the term $n_{{\rm tap},ij}\Delta tap_{ij}<<1$ and $v_{i}\approx v_{\rm nom}$ during normal operation. % and the second term is much less sensitive to $v_{\rm j}$ compared to $n_{\rm ij}$. 
The model errors of this linear approximation are presented in Fig. \ref{error} where the tap position ranges from $-16$ to $+16$. It can be seen that the model errors are no worse than $\pm 1$ tap position under a wide operation range and are no worse than $\pm 2$ tap position except some extreme cases that seldom happen in real operation, implying a sufficient accuracy used in our VVC problem. Note that, as discussed before, the impact of model errors can be compensated by the closed-loop nature of the lower-layer control to avoid severe voltage deviations.

Then, substituting (\ref{linearization}) into (\ref{BFM_alpha}c), one can obtain, $\forall(i,j)\in\mathcal{E}$,
\begin{align}\label{LGBFM}
\hspace{-3mm}v_{i}-v_{j}&=2(r_{ij}P_{ij}+x_{ij}Q_{ij})-2n_{{\rm tap},ij}\cdot\Delta{tap}_{ij}\cdot v_{{\rm nom}}.
\end{align}
Hence, the G-LBFM can be expressed by combining (\ref{BFM_alpha}a)-(\ref{BFM_alpha}b) and (\ref{LGBFM}).
\subsubsection{Extension to Unbalanced Three-Phase Systems}
\begin{figure}[t]
  \centering
  \includegraphics[width=2in]{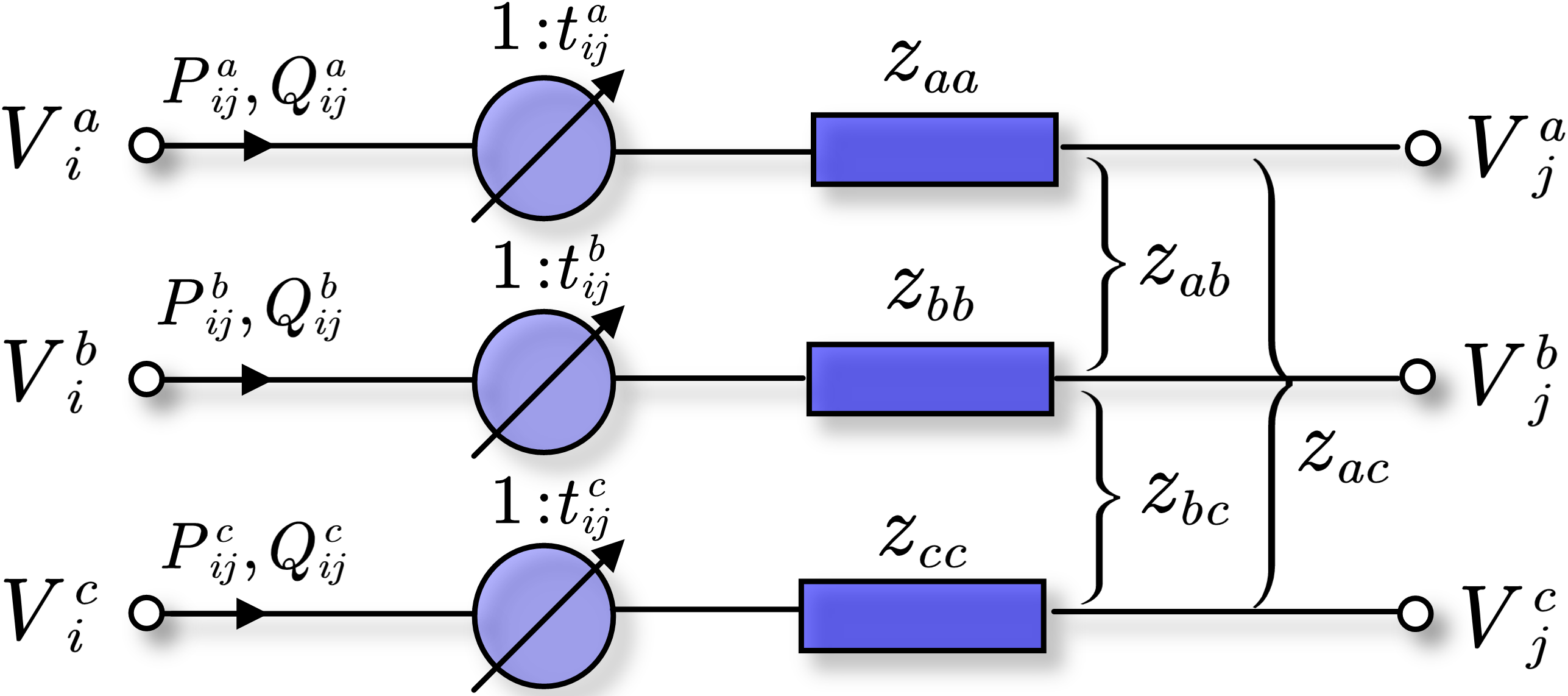}\\
  \caption{Generalized three-phase branch model.}\label{3ph_branch}
\end{figure}
The G-LBFM is extended to unbalanced three-phase systems (see Fig. \ref{3ph_branch}) where $z_{ij}$ is extended to ${\bm z}_{ij}:={\bm r}_{ij}+{\rm j}{\bm x}_{ij}\in\mathbb{C}^{3\times3}$, $n_{{\rm tap},ij}$ is extended to ${\bm n}_{{\rm tap},ij}:=[n_{{\rm tap},ij}^{\rm a},n_{{\rm tap},ij}^{\rm b},n_{{\rm tap},ij}^{\rm c}]^T$ and let $\Delta{\bm{tap}}_{ij}:=[\Delta{tap}_{ij}^{\rm a},\Delta{tap}_{ij}^{\rm b},\Delta{tap}_{ij}^{\rm c}]^T$. Correspondingly, the three-phase voltages, power consumption, power injections and line power flows in (\ref{BFM_alpha}) are extended to a three-phase form  ${\bm v}_{i}$, ${\bm p}_{{\rm c},j}$, ${\bm q}_{{\rm c},j}$, ${\bm p}_{{\rm inv},j}$, ${\bm q}_{{\rm inv},j}$, ${\bm q}_{{\rm cap},j}$, ${\bm P}_{ij}$ and ${\bm Q}_{ij}$, respectively.

%Then, based on the assumptions: i) bus voltage magnitudes on each phase are close, i.e. $|V_{i}^{\rm a}|\approx|V_{i}^{\rm b}|\approx|V_{i}^{\rm c}|$ and ii) phase unbalances on each bus are not too severe, i.e., $V_{i}^{\rm a}/V_{i}^{\rm b}\approx V_{i}^{\rm b}/V_{i}^{\rm c}\approx V_{i}^{\rm c}/V_{i}^{\rm a}\approx e^{\rm j2\pi/3}$, 

Assume the system unbalance is not too severe \cite{qianzhi1}, the G-LBFM is extended to a three-phase case as, $\forall (i,j)\in\mathcal{E}$,
\begin{subequations}\label{BFM_3ph}
\begin{align}
{\bm P}_{ij}&=\sum_{k\in\mathcal{N}_{j}}{\bm P}_{jk}+{\bm p}_{{\rm c},j}-{\bm p}_{{\rm inv},j}\\
{\bm Q}_{ij}&=\sum_{k\in\mathcal{N}_{j}}{\bm Q}_{jk}-{\bm q}_{{\rm c},j}-{\bm q}_{{\rm inv},j}-{\bm q}_{{\rm cap},j}\\
{\bm v}_{i}-{\bm v}_{j}&=2\left(\bar{\bm r}_{ij}{\bm P}_{ ij}+\bar{\bm x}_{ij}{\bm Q}_{ij}\right)+2{v}_{\rm nom}\cdot{\bm n}_{{\rm tap},ij}\odot\Delta{\bm{tap}}_{ij}
\end{align}
\end{subequations}
with
\begin{align*}
\bar{\bm r}_{ij}&={\rm Re}\{\bm{\alpha\alpha}^H\}\odot{\bm r}_{ij}+{\rm Im}\{\bm{\alpha\alpha}^H\}\odot{\bm x}_{ij},\\
\bar{\bm x}_{ij}&={\rm Re}\{\bm{\alpha\alpha}^H\}\odot{\bm x}_{ij}-{\rm Im}\{\bm{\alpha\alpha}^H\}\odot{\bm r}_{ij},
\end{align*}
where ${\bm\alpha}:=[1, e^{\rm-j2\pi/3}, e^{\rm j2\pi/3}]^T$ and $\odot$ denotes the element-wise multiplication.
\subsubsection{Compact Representation}
To better present the algorithm in the lower layer and some related analyses, we further rewrite the G-LBFM in a compact form. Firstly, similar as the compact single-phase model in \cite{haozhu}, the three-phase voltages (except bus 0), real/reactive power load and injections at bus and active/reactive power flows and tap position over each branch are compactly represented by column vectors ${\bm v}, {\bm p}_{\rm c}, {\bm p}_{\rm inv},{\bm q}_{\rm inv}, {\bm q}_{\rm cap}, {\bm q}_{\rm c}, \bm P, \bm Q$ and $\bm n$, respectively. Branch resistances and reactances are compactly represented by block-diagonal matrices $\bm R$ and $\bm X$, respectively. To be noticed, we only consider the buses and branches that actually exist in real systems instead of considering a completed three-phase system by adding some virtual buses and branches.
Let $\varphi_i$ and $\varphi_{ij}$ be the actual phase sets of bus $i$ and branch $(i,j)$. 
%Secondly, we remove the elements corresponding to the phase(s) which does not actually exist in real systems though it is uniformly represented in model (\ref{BFM_3ph}), so that some matrix properties will hold in the following analysis. For example, let $\varphi_{\rm ij}\subseteq\{\rm a,b,c\}$ be the phase set of branch $\rm (i,j)$. If ${\rm a}\notin\varphi_{\rm ij}$, then $P_{\rm ij}^{\rm a}, Q_{\rm ij}^{\rm a}$ should be removed from $\bf P, Q$ and the corresponding impedance entries from $\bf R$ and $\bf X$. 
Let $\bar{\bm G}:=[{\bm g}_0\,\,{\bm G}^T]^T\in\mathbb{R}^{(N+1)\times N}$ be the (single-phase) graph incidence matrix of the distribution network where ${\bm g}_0^T$ denotes the first row of $\bar{\bm G}$ \cite{incidencematrix}. The extended incidence matrix of a three-phase system $\bm{\bar{A}}:=[\bm{\bar{A}}_{ij}]_{\sum|\varphi_{ij}|\times\sum|\varphi_{ij}|}$ is defined as,
\begin{align}
\bar{\bm A}_{ij}=\bar{G}_{ij}\otimes{\bm I}_{|\varphi_{ij}|} 
\end{align}
where ${\bm I}_{|\varphi_{ij}|}$ is the $\varphi_i\times\varphi_j$ identity matrix; $j$ is the number index of branch $(i,j)$, i.e., $j{\rm th}$ column of $\bar{\bm G}$ corresponds to branch $(i,j)$.
Then, the compact G-LBFM of the whole system can be expressed as,
\begin{subequations}\label{CBFM}
\begin{align}
 \hspace{-3mm}-\bm{AP}&=-{\bm p}_{\rm inv}+{\bm p}_{\rm c}\\
\hspace{-3mm} -\bm{AQ}&=-{\bm q}_{\rm inv}-{\bm q}_{\rm cap}+{\bm q}_{\rm c}\\
    \hspace{-3mm} \begin{bmatrix}{\bm a}_0\,\,{\bm A}^T\end{bmatrix}\begin{bmatrix}{\bm v}_0\\{\bm v}\end{bmatrix}&=2(\bm{RP+XQ})+2{v}_{\rm nom}\cdot{\bm n}\odot\Delta\bm{tap}
\end{align}
\end{subequations}
where ${\bm a}_0^T$ denotes the submatrix of $\bar{\bm A}$ corresponding to bus $0$ while $\bm A$ denotes the remaining submatrix.

By substituting (\ref{CBFM}a) and (\ref{CBFM}b) into (\ref{CBFM}c), we have
\begin{align}\label{volt}
 {\bm v}=\bm{Mq}_{\rm inv}+\bm\mu
\end{align}
where
\begin{align*}
 &\hspace{25mm}{\bm M}=2{\bm A}^{-T}{\bm X}{\bm A}^{-1}\\
 {\bm\mu}=&{\bm A}^{-T}\Big(-{\bm a}_0{\bm v}_0+2{\bm R}{\bm A}^{-1}({\bm p}_{\rm inv}-{\bm p}_{\rm c})\\
 &\hspace{8mm}+2{\bm X}{\bm A}^{-1}\left({\bm q}_{\rm cap}-{\bm q}_{\rm c}\right)+2{v}_{\rm nom}\cdot{\bm n}\odot\Delta\bm{tap}\Big).
\end{align*}
Clearly, $\bm\mu$ can be considered as the component  contributed by all other factors except the VAR injections ${\bm q}_{\rm inv}$.

\subsection{Voltage Regulation Device Models}
\subsubsection{Step-Voltage Regulator} The SVRs can be modelled as an ideal transformer with a tap changer due to their small impedance. For single-phase SVRs, each phase has its own compensator circuit, so that the taps can be changed separately. For three-phase SVRs with only one compensator circuit, the taps on all windings change the same \cite{Kersting}. Here, both single-phase SVRs and three-phase SVRs are applied.
\subsubsection{OLTC Transformer}
For the substation OLTC transformer, by neglecting the shunt admittance, it can be modelled as a series impedance with an ideal three-phase voltage regulator, which can be also modelled by the G-LBFM.  
\subsubsection{Capacitor Bank}
The switchable CBs can be approximately modelled by constant reactive power sources with discrete characteristics. Similarly, they are connected in a single-phase way, which enables separable
VAR compensations on each phase. \iffalse The operation limits of SVRs can be expressed as, for any $\rm(i,j)\in\mathcal{E}_{SVR}$ and $\phi\in\{\rm a,b,c\}$,
\begin{align}
  \underline{n}_{\rm ij}^\phi\leq n_{\rm ij}^\phi\leq \overline{n}_{\rm ij}^\phi,\,\,n^\phi_{\rm ij}\in\mathbb{Z}
\end{align}\
\begin{align}
 q_{\rm CB}^{\phi}=n_{\rm CB}^{\phi}\Delta q_{\rm CB}^{\phi}
\end{align}
where $\underline{n}_{\rm ij}^\phi$ and $\overline{n}_{\rm ij}^\phi$ denotes the lower and upper bounds of tap position.\fi
\subsubsection{DER Inverter}
The voltage source inverter-based DER  has a typical  cascading control structure, i.e., inner current control loop and outer control loop, which enables the decoupled active and reactive power control. The total power outputs are limited by the inverter capacity and it is assumed that the active power has higher priority than reactive power when the total output power reaches to the capacity limit. Besides, in LV distribution systems, DER inverters are generally connected to a single phase, enabling separable VAR support on each phase.

\section{Receding Horizon Optimization for Optimal Scheduling of OLTC, SVRs and CBs}\label{sec:stage1}
In the upper-layer, a RHO-based VVC strategy is developed to deal with hourly voltage issues while improving the economic operation of distribution systems by optimally coordinating the operation of voltage regulation devices. Each term in the objective is formulated as follows,
\begin{itemize}
  \item Network power losses
   \begin{align}\label{losses}
        \hspace{-5mm}J_{\rm loss}(t):=\sum_{(i,j)\in\mathcal{E}}\sum_{\phi\in\varphi_{ij}}r_{ij}^{\phi\phi}\cdot\frac{\big(P_{ij}^\phi(t)\big)^2+\big(Q_{ij}^\phi(t)\big)^2}{v_{\rm nom}}.
    \end{align}
\item Operation costs of SVRs
\begin{align}\label{SVRs_change}
    \hspace{-5mm}J_{\rm tap}(t)&:=\sum_{(i,j)\in\mathcal{E}}\sum_{\phi\in\varphi_{ij}}\left(n_{{\rm tap},ij}^{\phi}(t)-n_{{\rm tap},ij}^{\phi}(t-1)\right)^2.
\end{align}
\item Operation costs of CBs
\begin{align}\label{CBs_change}
    \hspace{-5mm}J_{\rm cap}(t)&:=\sum_{i\in\mathcal{N}}\sum_{\phi\in\varphi_i}\left(N_{{\rm cap},i}^{\phi}(t)-N_{{\rm cap},i}^{\phi}(t-1)\right)^2.
\end{align}
\end{itemize}

Thus, the objective of the RHO-based VVC problem is formulated over a prediction length $T_H$ with the penalty factors $C_{\rm loss},C_{\rm loss},C_{\rm cap}$ and $C_\delta$ assigned to terms (\ref{losses})--(\ref{CBs_change}) and the voltage violation term.  
\begin{subequations}\label{VVC1}
\begin{align}
\nonumber&\text{minimize}\,\,\,\sum_{t=1}^{T_H}\Big(C_{\rm loss}\cdot J_{\rm loss}(t)+C_{\rm tap}\cdot J_{\rm tap}(t)\\
&\hspace{16mm}+C_{\rm cap}\cdot J_{\rm cap}(t)\Big)+C_\delta\cdot\sum_{i\in\mathcal{N}}\sum_{\phi\in\varphi_i}\big(\delta_i^\phi\big)^2\\
\nonumber&\text{over}\,\,\,n_{{\rm tap},ij}^{\phi}(t),N^{\phi}_{{\rm cap},i}(t),q_{{\rm inv},i}^{\phi}(t),P_{ij}^{\phi}(t),Q_{ij}^\phi(t),v_{i}^{\phi}(t),\delta_i^\phi,\\
\nonumber&\forall\,t\in\{1,\ldots,T_H\},\,\forall\,({i,j})\in\mathcal{E},\,\forall\, i,j\in\mathcal{N},\,\forall\,\phi\in\{\rm a,b,c\}\\
&\nonumber\text{s}\text{ubject}\,\,\text{to}\\
&P_{ij}^{\phi}(t)=\sum_{k\in\mathcal{N}_{j}}P_{jk}^{\phi}(t)+{\hat p}_{{\rm c},j}^{\phi}(t)-{\hat p}_{{\rm inv}, j}^{\phi}(t)\\
&Q_{ij}^{\phi}(t)=\sum_{k\in\mathcal{N}_{j}}Q_{jk}^{\phi}(t)+{\hat q}_{{\rm c},j}^{\phi}(t)-q_{{\rm inv}, j}^{\phi}(t)-q_{{\rm cap},j}^\phi(t)\\
&\nonumber v_{i}^{\phi}(t)-v^{\phi}_{j}(t)=\sum_{\phi^\prime\in\varphi_{ij}}2\left(\overline{r}^{\phi\phi^\prime}_{ij}P^{\phi}_{ij}(t)+\overline{x}^{\phi\phi^\prime}_{ij}Q^{\phi}_{ij}(t)\right)\\
&\hspace{25mm}+2{v}_{\rm nom}\cdot n_{{\rm tap},ij}^\phi(t)\cdot\Delta{tap}^\phi_{ij}\\
&\underline{n}_{{\rm tap},ij}^{\phi}\leq{n}_{{\rm tap},ij}^{\phi}(t)\leq\overline{n}_{{\rm tap},ij}^{\phi},\,\,{n}_{{\rm tap},ij}^{\phi}(t)\in\mathbb{Z}\\
&\left|{n}_{{\rm tap},ij}^{\phi}(t)-{n}_{{\rm tap}, ij}^{\phi}(t-1)\right|\leq\Delta\overline{n}_{{\rm tap},ij}^{\phi}\\
&\sum_{i=1}^T\left|{n}_{{\rm tap},ij}^{\phi}(t)-{n}_{{\rm tap},ij}^{\phi}(t-1)\right|\leq\Delta_\Sigma\overline{n}_{{\rm tap},ij}^{\phi}\\
&q_{{\rm cap},i}^\phi(t)=N_{{\rm cap},i}^{\phi}(t)\cdot\Delta q_{{\rm cap},i}^{\phi},\,\,{N}_{{\rm cap},i}^{\phi}(t)\in\mathbb{Z}\\
&0\leq{N}_{{\rm cap},i}^{\phi}(t)\leq\overline{N}_{{\rm cap},i}^{\phi}\\
&\left|{N}_{{\rm cap},i}^{\phi}(t)-{N}_{{\rm cap},i}^{\phi}(t-1)\right|\leq\Delta\overline{N}_{{\rm cap},i}^{\phi}\\
&\sum_{t=1}^{T}\left|{N}_{{\rm cap},i}^{\phi}(t)-{N}_{{\rm cap},i}^{\phi}(t-1)\right|\leq\Delta_\Sigma\overline{N}_{{\rm cap},i}^{\phi}\\
&\underline{q}_{{\rm inv},i}^\phi(t)\leq q_{{\rm inv},i}^\phi(t)\leq\overline{q}_{{\rm inv},i}^\phi(t)\\
&\hspace{-1mm}-\underline{q}_{{\rm inv},i}^{\phi}(t)=\overline{q}_{{\rm inv},i}^{\phi}(t)=\eta\sqrt{(s_{{\rm inv},i}^{\phi})^2-(p_{{\rm inv},i}^\phi(t))^2}\\
&\underline{v}_{i}^{\phi}-\delta_i^\phi\leq v_{i}^{\phi}(t)\leq\overline{v}_{i}^{\phi}+\delta_i^\phi.
\end{align}
\end{subequations}
The constraints (\ref{VVC1}b)--(\ref{VVC1}d) represent the power flow constraints where ${\hat p}_{{\rm c},j}^{\phi}, {\hat p}_{{\rm inv},j}^\phi$ and ${\hat q}_{{\rm inv},j}^\phi$ can be obtained from real-time prediction. (\ref{VVC1}e)--(\ref{VVC1}k) denote the operation limits of SVRs and CBs, respectively.\footnote{Let $\overline{N}_{{\rm cap},i}^{\phi}=0$ for those buses/phases without CBs and $\underline{n}_{{\rm tap},ij}^{\phi}=\overline{n}_{{\rm tap},ij}^{\phi}=0$ for branches without SVRs.} (\ref{VVC1}l)-(\ref{VVC1}m) denote the operation constraints of DER inverters. In this layer, we reserve partial VAR capabilities of inverters in the optimization by introducing a scalar $0<\eta<1$, in which way, the effects of prediction errors of DERs can be effectively reduced. Otherwise, it might result in conservative operations of other discrete devices and, consequently, the lower-layer control might fail to regulate the voltages within the predefined range.
(\ref{VVC1}n) denotes the voltage constraints. To avoid infeasible cases caused by the voltage limit, the slack term $\delta^{\phi}_i$ is introduced which will be significantly punished in the objective function as the last term.
For the branch $(i,j)$  with a three-phase SVR with only one compensator circuit, an additional constraint on tap changers should be added into (\ref{VVC1}), which is as follows,
\begin{align}
 n_{{\rm tap},ij}^a(t)=n_{{\rm tap},ij}^b(t)=n_{{\rm tap},ij}^c(t),\,\,\forall t.
\end{align}
To be noticed, given the constraints in (\ref{VVC1}g) and (\ref{VVC1}k) with a sum of absolute terms cannot be directly handled by solvers, auxiliary variables $n_{{\rm tap},ij,+}^{\phi}(t),n_{{\rm tap},ij,-}^{\phi}(t), N_{{\rm cap},ij,+}^{\phi}(t)$ and $ N_{{\rm cap},ij,-}^{\phi}(t)$ are introduced to transform the constraints into an equivalent linear form. Constraint (\ref{VVC1}g) is equivalently expressed by,
\begin{subequations}
\begin{align}
&\hspace{-2mm}\sum_{t=1}^{T}\left(n_{{\rm tap},ij,+}^{\phi}(t)+n_{{\rm tap},ij,-}^{\phi}(t)\right)\leq\Delta_\Sigma\overline{n}_{{\rm tap},ij}^{\phi}\\
&\hspace{-2mm}n_{{\rm tap},ij,+}^{\phi}(t)-n_{{\rm tap},ij,-}^{\phi}(t)=n_{{\rm tap},ij}^{\phi}(t)-n_{{\rm tap},ij}^{\phi}(t-1)\\[1mm]
&\hspace{-2mm}n_{{\rm tap},ij,+}^{\phi}(t)\geq0,\,\, n_{{\rm tap},ij,-}^{\phi}(t)\geq0
\end{align}
\end{subequations}
and, similarly, (\ref{VVC1}k) becomes
\begin{subequations}\label{CBreformulate}
\begin{align}
&\hspace{-1mm}\sum_{t=1}^{T}\left(N_{{\rm cap},i,+}^{\phi}(t)+N_{{\rm cap},i,-}^{\phi}(t)\right)\leq\Delta_\Sigma\overline{N}_{{\rm cap},i}^{\phi}\\
&\hspace{-1mm}N_{{\rm cap},i,+}^{\phi}(t)-N_{{\rm cap},i,-}^{\phi}(t)=N_{{\rm cap},i}^{\phi}(t)-N_{{\rm cap},i}^{\phi}(t-1)\\[1mm]
&\hspace{-1mm}N_{{\rm cap},i,+}^{\phi}(t)\geq0, N_{{\rm cap},i,-}^{\phi}(t)\geq0.
\end{align}
\end{subequations}
After such reformulations, the problem (\ref{VVC1}) becomes a  (convex) mixed-integer quadratic programming (MIQP) problem  and can thus be efficiently solved by MIQP solvers. Only the  first-step optimal solution  will be used to schedule the OLTC transformer, SVRs and CBs while the corresponding voltage solution will be used in the lower-layer control.

\section{Decentralized Integral-Like Control for Reactive Power Dispatch of DER Inverters}\label{sec:stage2}
In this layer, a decentralized control scheme for VAR dispatch of DER inverters is developed to track the voltage references scheduled by the upper-layer control and mitigate the voltage fluctuations in the timescale of seconds.
\subsection{Integral-Like Algorithm}
Firstly, the (squared) voltage reference at bus $(i,\phi)$ is scheduled by\footnote{Here, we abuse the notation of time stamp $t$ in the upper and lower layers, which actually corresponds to different time intervals.},
\begin{align}\label{vref}
  {v}_{{\rm ref},{i}}^\phi(t)=\left[\nu_i^{\phi}\right]^{\overline{v}_{i}^{\phi}}_{\underline{v}_{i}^{\phi}},\, \phi\in\varphi_i
\end{align}
where $\nu_i^{\phi}$ denotes the first-step optimal voltage solution at bus $(i,\phi)$ in the RHO-based upper-layer control; $[\ast]^{\overline{v}_{i}^{\phi}}_{\underline{v}_{i}^{\phi}}$ denotes the projection operator onto the constraint set $[{\underline{v}_{i}^{\phi}},{\overline{v}_{i}^{\phi}}]$. The projection is necessary here because we introduce a slack variable in the voltage constraint in the upper-layer optimization, which implies there might be voltage solution on some buses violating the predefined limit. The voltage reference (\ref{vref}) will be maintained unchanged during every upper-layer control period.

Each DER inverter connected to bus $(i,\phi)$ updates its VAR output according to the integral-like control law,
\begin{align}\label{DeVVC}
\hspace{-2mm}{q}_{{\rm inv},i}^{\phi}(t+1)=\left[{q}_{{\rm inv},i}^{\phi}(t)-\gamma\left({v}_{i}^{\phi}(t)-{v}_{{\rm ref},{i}}^\phi(t)\right)\right]^{\overline{q}_{{\rm inv},i}^\phi(t)}_{\underline{q}_{{\rm inv},i}^\phi(t)}
\end{align}
or compactly,
\begin{align}
 \hspace{-2mm}{\bm q}_{{\rm inv}}(t+1)=\left[{\bm q}_{{\rm inv}}(t)-\gamma\left({\bm v}(t)-{\bm v}_{{\rm ref}(t)}\right)\right]^{\overline{\bm q}_{{\rm inv}}(t)}_{\underline{\bm q}_{{\rm inv}}(t)}
\end{align}
where $\gamma>0$ denotes the step size; ${\overline{q}_{{\rm inv},i}^\phi(t)}=-{\underline{q}_{{\rm inv},i}^\phi(t)}=\sqrt{(s_{{\rm inv},i}^{\phi})^2-(p_{{\rm inv},i}^\phi(t))^2}$. 
Accordingly, each DER inverter updates its VAR output according to its local instantaneous phase voltage measurement.
\subsection{Stability Analysis}
The closed-loop system stability (the distribution network with the decentralized VAR controllers) under a fixed point $(\bm v_{\rm ref}, \underline{\bm q}_{\rm inv}, \overline{\bm q}_{\rm inv})$ is investigated. Let $({\bm q}_{\rm inv}^\ast,{\bm v}^{\ast})$ be the stationary point of closed-loop system. According to ($\ref{DeVVC}$), we have
\begin{align}
\nonumber&\left\|{\bm q}_{\rm inv}(t+1)-{\bm q}_{\rm inv}^\ast\right\|\\
\nonumber&\hspace{0mm}=\left\|\left[{\bm q}_{\rm inv}(t)-\gamma\left({\bm v}-{\bm v}_{\rm ref}\right)\right]^{\overline{\bm q}_{\rm inv}}_{\underline{\bm q}_{\rm inv}}-\left[{\bm q}_{\rm inv}^\ast-\gamma\left({\bm v}^\ast-{\bm v}_{\rm ref}\right)\right]^{\overline{\bm q}_{\rm inv}}_{\underline{\bm q}_{\rm inv}}\right\|\\
\nonumber&\hspace{0mm}\leq\left\|{\bm q}_{\rm inv}(t)-\gamma\left({\bm v}(t)-{\bm v}_{\rm ref}\right)-{\bm q}_{\rm inv}^\ast+\gamma\left({\bm v}^\ast-{\bm v}_{\rm ref}\right)\right\|\\
\nonumber&\hspace{0mm}=\left\|{\bm q}_{\rm inv}(t)-\gamma\left({\bm M}\bm q_{\rm inv}(t)+\bm\mu-{\bm v}_{\rm ref}\right)\right.\\
\nonumber&\hspace{38mm}\left.\ -{\bm q}_{\rm inv}^\ast+\gamma\left({\bm M}\bm q_{\rm inv}^\ast+\bm\mu-{\bm v}_{\rm ref}\right)\right\|\\
&\hspace{0mm}\leq\left\|{\bm I}-\gamma{\bm M}\right\|\left\|{\bm q}_{\rm inv}(t)-{\bm q}_{\rm inv}^\ast\right\|
\end{align}
where the first inequality holds because the projection operation is a non-expensive mapping. Accordingly, a sufficient stability condition that guarantees $\left\|{\bm q}_{\rm inv}(t)-{\bm q}_{\rm inv}^\ast\right\|$ contracting at each iteration is 
\begin{align}\label{stabilitycond}
\left\|\bm I-\gamma\bm M\right\|<1
\end{align}
which provides a upper bound of $\gamma$ to avoid hunting between independent VAR control loops of DERs.
\section{Case Study}\label{sec:case}
\begin{figure}[t]
  \centering
  \includegraphics[width=3.42in]{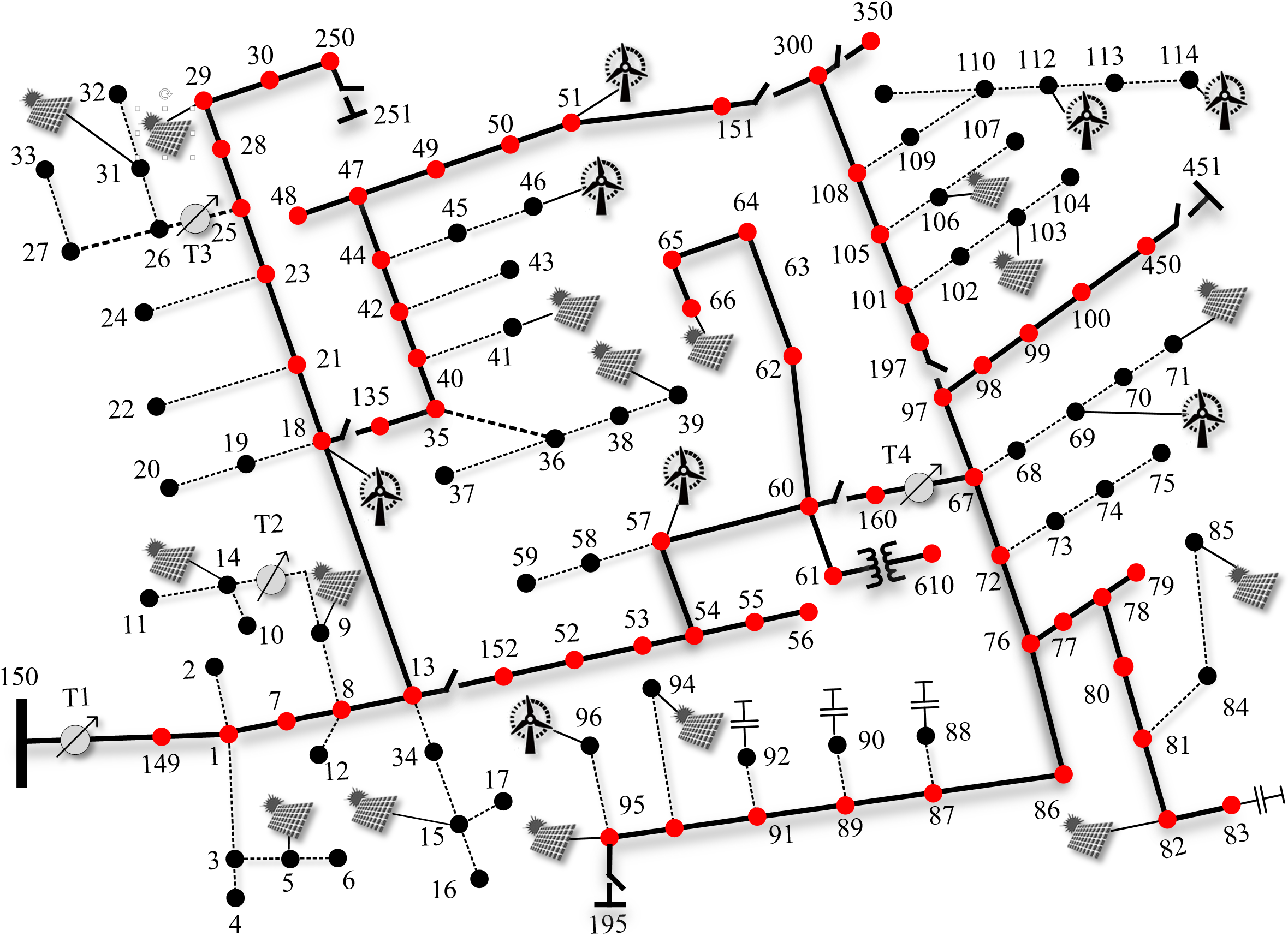}\\
  \caption{IEEE 123-Node Test Feeder.}\label{IEEE123}
\end{figure}
In this section, the proposed VVC framework is verified on the unbalanced IEEE 123-Node Test Feeder \cite{testfeeder}. Multiple PV and full-inverter wind turbines (WTs), each with a rated capacity of 200 kW, are integrated into the system with the locations shown in Fig. \ref{IEEE123}.
We develop a simulation framework in MATLAB R2019b, which integrates YALMIP Toolbox \cite{yalmip} with IBM ILOG CPLEX 12.9 solver \cite{cplex} for optimization, and the open-source Open Distribution System Simulator (OpenDSS) \cite{opendss} for power flow analysis. The OpenDSS can be controlled from MATLAB through a component object model interface, allowing us to carry out the VVC algorithms, perform power flow calculations, and retrieve the results. The control mode is disabled in OpenDSS to ensure that the tap changers are not automatically adjusted according to  local voltages during a simulation.
The daily residential load and solar profiles are obtained from the historical data \cite{Holcomb} and the wind power time-series is generated from the wind datasets provided by National Renewable Energy Laboratory \cite{winddata}.

%The simulations are carried out using the open-source simulator OpenDSS \cite{opendss} based on the full nonlinear ac power flow model instead of the linear approximation, which is interfaced with MATLAB R2019b to perform the control algorithms implemented with the YALMIP Toolbox \cite{yalmip} and the solver IBM ILOG CPLEX 12.9. All the simulations are run on a PC with an Intel Xeon CPU E3-1240-v5 @3.50 GHz processor and 16 GB RAM. 
%The static and dynamic performances of the proposed methods will be presented and compared with several existing methods.
\subsection{Static Performance}
\subsubsection{Upper-Layer Control}
\begin{table}
\centering
\caption{Solution Quality With Different Models}\label{ULC_static1}
\footnotesize
\renewcommand\arraystretch{1.0}
  %\centering\caption{Comparison Between Distributed and Centralized Control}\label{}
  %\renewcommand{\multirowsetup}{\centering}
        \begin{tabular}{ccc}
        \hline
        \hline
         &{Linear Model}&{ Nonlinear Model}\\
         \hline
        \specialrule{0.0em}{0mm}{0.5mm}
        Losses& 24.14 kW& 118.67 kW\\[0.3mm]
        \hline
        \hline
        \end{tabular}
\end{table}
\begin{table}[t]
\centering
\caption{Computation Time Under Different Models}\label{ULC_static2}
\footnotesize
\renewcommand\arraystretch{1.0}
  %\centering\caption{Comparison Between Distributed and Centralized Control}\label{}
  %\renewcommand{\multirowsetup}{\centering}
        \begin{tabular}{ccccc}
        \hline
        \hline
         $T_H$&{Linear Model}&{ Nonlinear Model}\\
         \hline
      \specialrule{0.0em}{0mm}{0.5mm}
        1&0.5045 s &5.625 s\\[0.5mm]
        3&0.8666 s&15.11 s\\[0.5mm]
        5&1.294 s&760.5 s\\[0.5mm]
        10&15.51 s&151.4 s\\[0.5mm]
        12&181.5 s&382.5 s\\[0.5mm]
        \hline
        \hline
        \end{tabular}
\end{table}
The computation performance of the upper-layer control under a static case is shown here. We take $T_H=1$, $C_{\rm tap}=C_{\rm cap}=0$ and $\eta=0.8$ to compare the quality of the solutions. The optimization problem with the G-LBFM constraint can be directly solved by CPLEX 12.9 while the problem with original nonlinear GBFM constraint is solved by IPOPT 3.13. Besides, we further make a comparison with the continuous relaxation-based method \cite{BAR_dis_tap}, wherein the relaxed discrete variables are rounded to the closest integer values.
The solutions are used to schedule the VVC devices in the test system (in OpenDSS) and the resultant real losses are measured, which are shown in Table \ref{ULC_static1}. It can be seen that the proposed method with the linear approximation is able to compute a much better solution than the one obtained  by continuous relaxation-based method. We further compare the computation efficiency of the two methods under different receding horizon, which is illustrated in Table \ref{ULC_static2}. Obviously, the proposed method shows much better computation efficiency, which is beneficial for the real-time control.

\subsubsection{Lower-Layer Control}
\begin{figure}[t]
  \centering
  \includegraphics[width=3.2in]{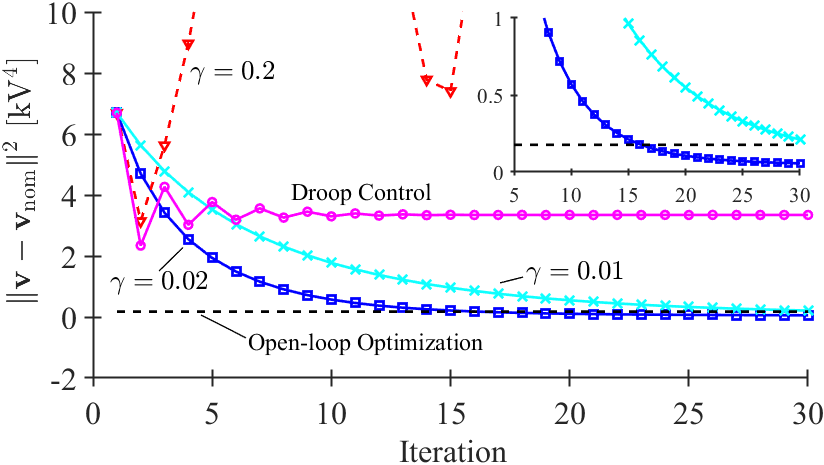}\\
  \caption{Static performance with different methods for lower-layer control.}\label{static_LLC}
\end{figure}

The tracking performance of the lower-layer control algorithm under a static case is also tested here to illustrate the convergence performance. We compare our methods\footnote{Note that, in the test of our method (also for the later droop control), at each iteration, the real voltage measurements (from OpenDSS) are used as feedback to perform the update instead of using the computation results from the linearized model.} with the droop control\footnote{We originally select the droop gain according to the widely-used principle \cite{Czhang_multi_time}, but due to the large gain, the droop control fails to converge under our problem, i.e., the closed-loop system is unstable, which is not shown in Fig. \ref{static_LLC}. Therefore, we gradually reduce the gain until it converges and the corresponding result is illustrated in the figure.} and the open-loop optimization method\footnote{The optimization method is used to minimize $\left\|\bm v-\bm v_{\rm ref}\right\|^2$ over the constraints $\bm q\in[\underline{\bm q}_{\rm inv},\overline{\bm q}_{\rm inv}]$ and linearized BFM (so that the problem is convex) using the solver CPLEX 12.9.}. The sufficient stability condition calculated from (\ref{stabilitycond}) is $\gamma<0.022$.
Fig. \ref{static_LLC} shows the convergence performance of different methods with the metric $\|\bm v-\bm v_{\rm ref}\|^2$. It can be seen that the proposed integral-like method with $\gamma=0.01$ or $0.02$ can efficiently converge to a stationary point within 20 iterations, implying a good tracking capability. In comparison, the droop control converges to an equilibrium after several steps of oscillation which is remote to the voltage reference. This is because the droop control is essentially the difference control, however, the proposed integral-like method is the non-difference control. The final result of the proposed method (see the zoomed part) is even better than that of the optimization-based method due to the online feedback mechanism. Additionally, a larger $\gamma$ naturally implies faster response and thus better tracking capability. But as $\gamma$ increases, the closed-loop system tends to be oscillating and finally diverges. So, there will be a trade off between fast tracking and stability in real-life applications.

\subsection{Dynamic Simulation}
\begin{figure}[t]
  \centering
  \includegraphics[width=3.25in]{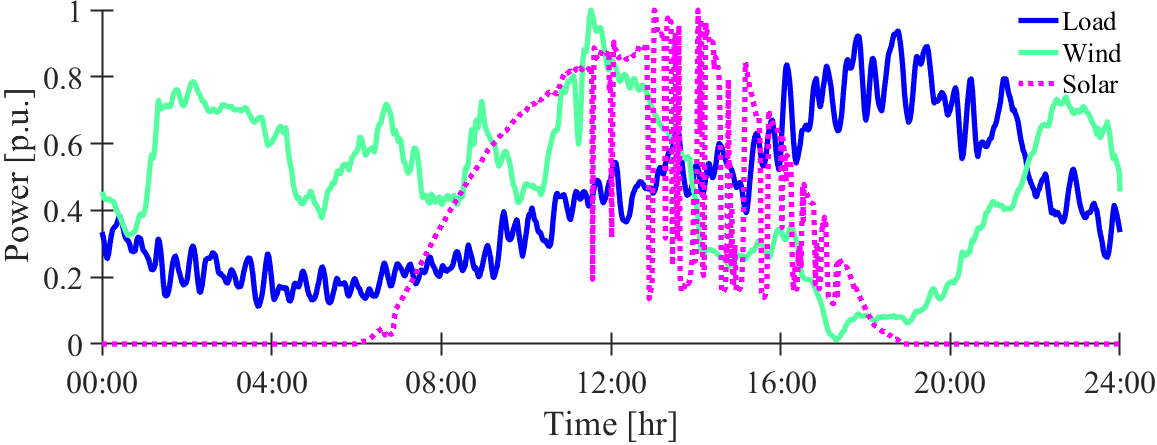}\\
  \caption{Daily residential load, wind and solar power profiles.}\label{timeseries}
\end{figure}
\begin{figure}[t]
  \centering
  \includegraphics[width=3.2in]{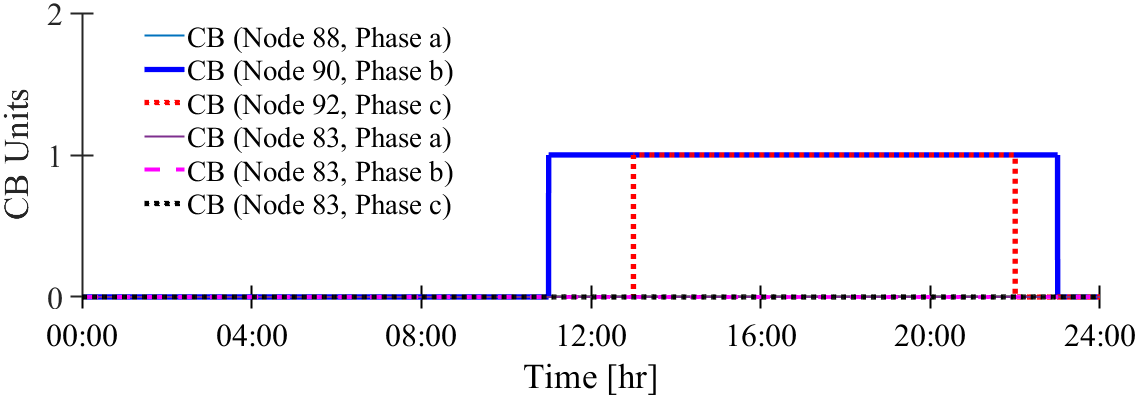}\\
  \caption{Operation of CBs.}\label{CB}
\end{figure}
\begin{figure}[t]
  \centering
  \includegraphics[width=3.1in]{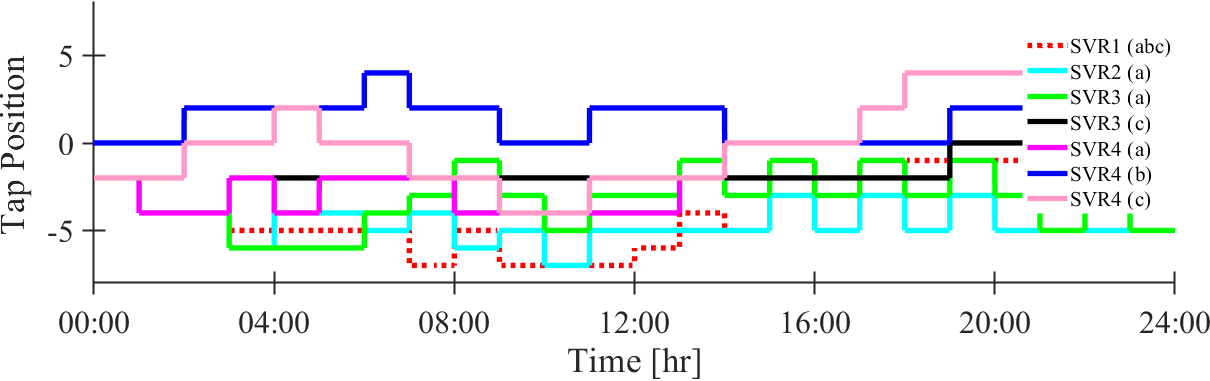}\\
  \caption{Operation of tap changers.}\label{TAP}
\end{figure}
\begin{figure}[t]
  \centering
  \includegraphics[width=3.1in]{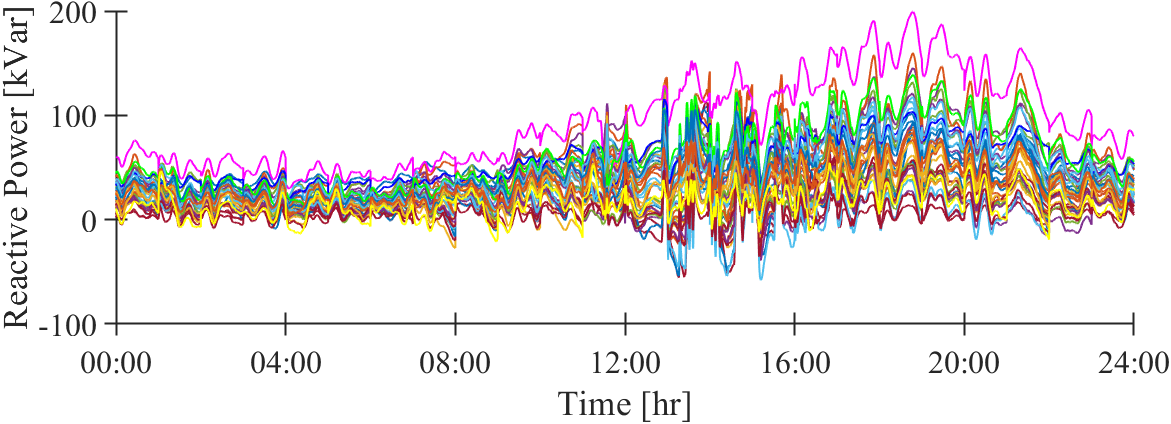}\\
  \caption{Reactive power outputs of DER inverters.}\label{qinv}
\end{figure}

The dynamic simulation is also performed. The daily load, solar and wind power profile are illustrated in Fig. \ref{timeseries}. In the proposed method, the upper-layer control period is 1 h and in the lower layer, VAR outputs of DER inverters are updated every 5 s. Besides, we set $T_H=3$ and $\gamma=0.02$. To better demonstrate the effectiveness of the proposed method, it is compared with cases with no control, with only upper-layer control, as well as with the continuous-relaxation-based control method. The operation of tap changers and CBs, as well as the VAR outputs of DERs over 24 hours are shown in Fig. 
\ref{CB}--\ref{qinv}, respectively.

\begin{figure}[h!]
  \centering
  \includegraphics[width=3.1in]{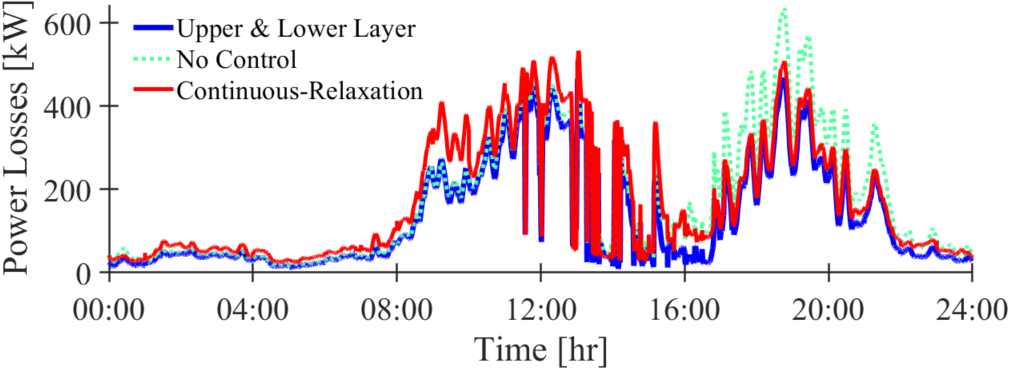}\\
  \caption{Power losses.}\label{loss}
\end{figure}
\begin{figure}[h!]
\centering
\subfloat[]{
\includegraphics[width=3.1in]{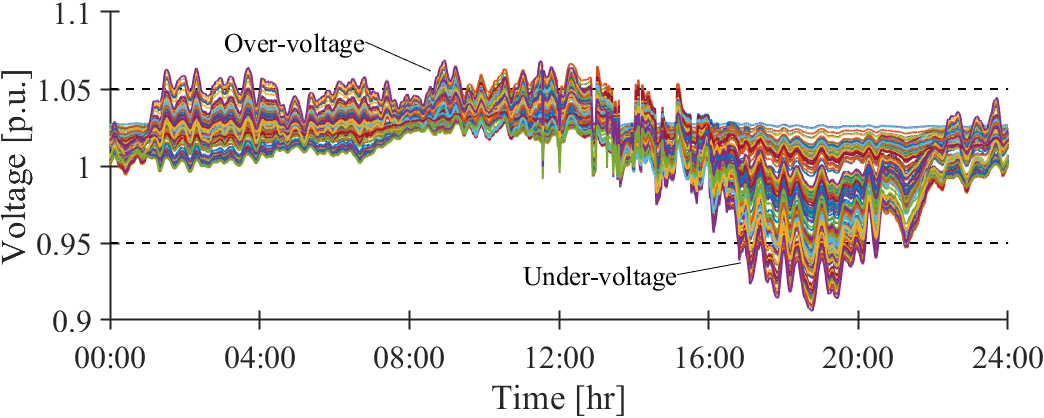}
}
\vspace{-1mm}
\subfloat[]{
\includegraphics[width=3.1in]{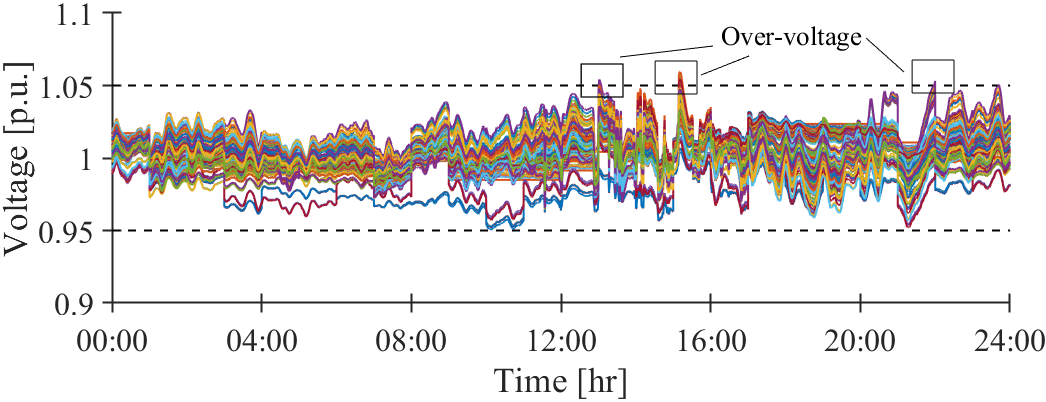}
}
\vspace{-1mm}
\subfloat[]{
\includegraphics[width=3.1in]{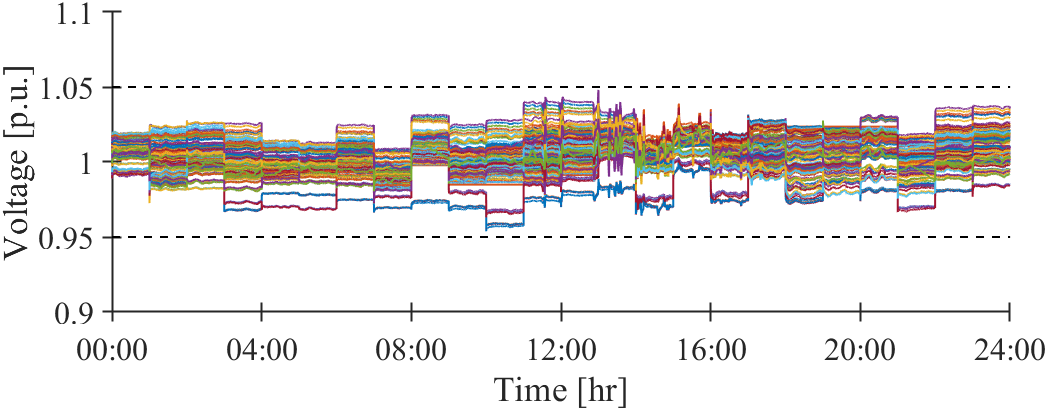}
}
\caption{Voltage profiles with different control strategies: (a) no control, (b) upper-layer control, and (c) the proposed two-layer control. }
\label{Vprofile}
\end{figure}
\begin{figure}[h!]
  \centering
 \includegraphics[width=3.15in]{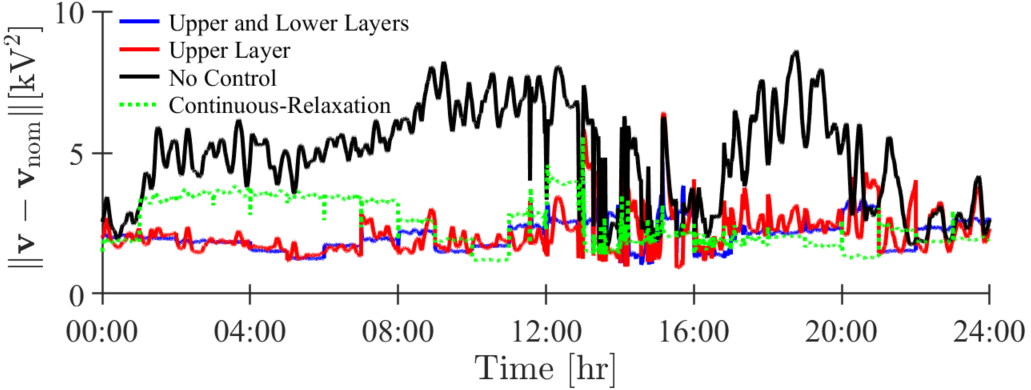}\\
 \caption{Overall voltage deviations with different strategies.}\label{dV}
\end{figure}
\begin{figure}[h!]
  \centering
 \includegraphics[width=3.15in]{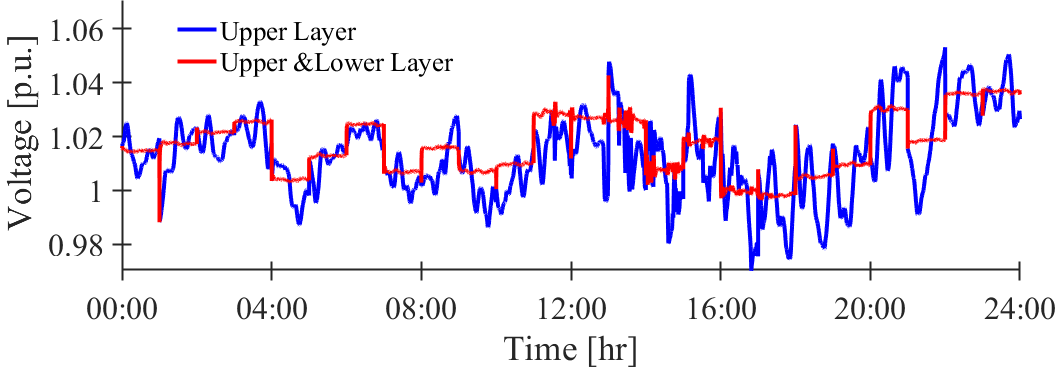}\\
 \caption{Voltage of Phase a at Node 114 with and without lower-layer control.}\label{V114}
\end{figure}

Fig. \ref{loss} shows the total power losses of the system. It can be seen that the proposed method with two-layer control can effectively reduce the losses, especially during 16:00--24:00 with heavy load consumption and show much better performance than the continuous-based method. The voltage profiles with different control strategies are illustrated in Fig. \ref{Vprofile}. Without Volt/VAR control, the voltages violate the range $[0.95,1.05]$ p.u. when DERs have high production e.g., 00:00--04:00 and 09:00--13:00 or when the load demand is heavy, e.g., 16:00--20:00. Only With the upper-layer control, in most time, the voltages can be regulated  within the predefined range $[0.95,1.05]$ p.u. with the help of tap changers and CBs, which validates the effectiveness of the open-loop upper-layer control. However, there are still some bus voltages that exceed $1.05$ p.u. in some moments (highlighted in Fig. \ref{Vprofile}(b)) due to the high fluctuations of DERs and load. 
In contrast, the two-layer control method can effectively regulate the voltages within the feasible range all the time, and due to the VAR support from DERs, the fast voltage fluctuations are significantly reduced, which can also be clearly observed from the overall voltage performance shown in Fig. \ref{dV} and the representative bus voltage at Node 114 (Phase a) shown in Fig. \ref{V114}, located in the end of the feeder. It can be observed that the voltage under the two-layer control can accurately track the reference, validating the good tracking capability of the  integral-like control algorithm in lower layer. Besides, given the control parameter $\gamma$ is designed within the stability boundary, there is no Volt/VAR hunting phenomenon.


\section{Conclusion}\label{sec:con}
This paper proposed a two-layer VVC framework for unbalanced distribution systems with high-penetration of inverter-based DERs. The upper layer optimally coordinates the OLTC, SVRs and CBs to improve economic benefits while correcting the long-term voltage deviations based on the RHO method. A G-LBFM was proposed to model the three-phase system with tap changers so that the optimization problem can be efficiently solved in real time. A decentralized integral-like algorithm was developed in the lower layer to deal with the fast voltage fluctuations by exploiting the VAR capabilities of DER inverters. Compared with the cases of no control and only upper layer control, the simulation results show that the proposed two-layer VVC method can effectively reduce the system losses while regulating the voltages within a predefined range in different timescales, due to the combination of open-loop optimization and closed-loop voltage reference tracking. Compared with the continuous-relaxation-based method, the proposed optimization model in upper layer can provide a much better solution and show much better computation efficiency. Compared with the droop-based real-time control, the proposed integral-like control in lower layer can better reduce the voltage deviations by fully exploiting the VAR capabilities of inverter-based DERs. The lower layer control also shows better numerical results than the optimization-based method due to its closed-loop nature.

\ifCLASSOPTIONcaptionsoff
  \newpage
\fi


\end{document}